\documentclass[11pt]{article}

\usepackage[final]{acl}

\usepackage{times}
\usepackage{latexsym}

\usepackage[T1]{fontenc}
\usepackage[utf8]{inputenc}

\usepackage{microtype}

\usepackage{inconsolata}

\usepackage{graphicx}

\title{Intellectual Humility as a Cognitive Filter for AI-Generated Health Misinformation. An Evolutionary Perspective on Epistemic Vigilance}

\author{
  \textbf{Marcin Rządeczka\textsuperscript{1}},
  \textbf{Maciej Wodziński\textsuperscript{1}},
  \textbf{Kacper Zacharski\textsuperscript{1}} and
  \textbf{Marcin Moskalewicz\textsuperscript{1,2,3}}
\\
  \textsuperscript{1}Maria Curie-Skłodowska University, Plac Marii Curie-Skłodowskiej 4, 20-031 Lublin, Poland\\
  \textsuperscript{2}IDEAS Research Institute, Krakowskie Przedmieście 13,
00-071 Warszawa, Poland\\
  \textsuperscript{3}Poznan University of Medical Sciences, Collegium Maius, Fredry 10, 61-701 Poznań, Poland\\
  \small{
    \textbf{Correspondence:} \href{mailto:email@domain}{marcin.rzadeczka@umcs.pl}
  }
}

\begin{document}
\maketitle
\begin{abstract}
We present experimental findings from a study (N=99) examining how intellectual humility (IH), i.e., the metacognitive awareness of one's epistemic limitations, affects the evaluation of AI-generated health dialogues varying in scientific rigor. Participants were randomly assigned to evaluate one of three dialogues about exercise and mental health: scientifically accurate, moderately pseudoscientific, or strongly pseudoscientific. Results reveal that IH functions as a selective cognitive filter. Individuals with higher humility scores rated pseudoscientific content as significantly less credible, while showing no correlation with credibility assessments of accurate content. Crucially, humility did not predict the ability to identify AI as the source of dialogues, suggesting that epistemic vigilance operates on content quality rather than source attribution. We interpret these findings through an evolutionary lens, proposing that IH represents an ancestral adaptation for navigating informationally uncertain environments. It remains effective at detecting exploitation attempts in AI-generated content, despite humans lacking evolved mechanisms for detecting AI sources. The study contributes to understanding how foundation models might improve or undermine human epistemic defenses, especially in health communication contexts.
\end{abstract}

\section{Introduction}

The proliferation of large language models (LLMs) capable of generating fluent and persuasive text has introduced unprecedented challenges for human epistemic vigilance, i.e., our evolved capacity to evaluate the reliability of communicated information \cite{Sperber2010}. Health misinformation, presenting risks as false beliefs about treatments or preventive measures, can directly impact physical well-being \cite{SwireThompson2020}. However, the psychological mechanisms underlying vulnerability or resistance to AI-generated health misinformation still remain poorly understood. 

The paper addresses a fundamental question at the intersection of cognitive science and ethical and responsible AI: do individual differences in intellectual humility (IH), defined as recognition of one's epistemic limitations and openness to revising beliefs \cite{Leary2017}, predict resistance to AI-generated pseudoscientific health claims? We treat IH not merely as a personality trait but also as a phenotypic manifestation of ancestral adaptations for managing epistemic uncertainty in complex social environments.

Our study makes three contributions relevant to the study of foundation models for social good. Firstly, we provide empirical evidence that IH selectively filters pseudo-scientific content while preserving openness to accurate information, which is a pattern consistent with error management theory \cite{Haselton2000} rather than general skepticism. Secondly, we demonstrate that source attribution (human vs. AI) operates independently of content evaluation, suggesting that interventions targeting AI detection may be less effective than those strengthening content-based epistemic vigilance. Thirdly, we offer an evolutionary framework for understanding why certain cognitive traits may provide natural defenses against AI-generated misinformation, suggesting the responsible deployment of foundation models in health communication contexts.

\section{Related Work}

\subsection{Intellectual Humility and Misinformation}

IH has recently emerged as a significant conceptual construct in misinformation research. \cite{KrumreiMancuso2015} developed the Comprehensive IH Scale and linked the trait to lower susceptibility to conspiracy theories and greater reliance on expert consensus. \cite{Leary2017} operationalized the construct as encompassing four dimensions: independence of intellect, openness to revising viewpoints, respect for different perspectives, and lack of intellectual overconfidence. Subsequent work has demonstrated that epistemically humble individuals show greater discernment between true and false claims across various domains \cite{Koetke2021}. \cite{Bak2022} provides a comprehensive review, positioning IH as an epistemic virtue with deep philosophical roots that have only recently become subject to empirical psychological investigation, and identifies four research strands: personality traits, cognitive functioning, social relations, and religiosity. Our study extends their cognitive functioning strand to a novel context, i.e., AI-generated misinformation, demonstrating that IH operates as a selective filter against pseudoscience rather than inducing general skepticism, while adding an evolutionary perspective absent from the current literature. Recently, \cite{Porter2024} found that IH predicted greater engagement with opposing political views and reduced polarization. 

However, existing work has focused primarily on human-authored misinformation. Notably, research has not examined whether the protective effects of IH extend to misinformation generated by artificial agents, where stylistic markers of unreliability may differ substantially from human-created content. The present study extends this literature to AI-generated health content, a domain where the stakes of false beliefs are particularly high. Furthermore, while previous studies have treated IH as a unitary protective factor, our design allows examination of whether humility functions selectively, i.e., by filtering pseudoscientific content while preserving receptivity to accurate information.

\subsection{AI Detection and Source Attribution}

Research on human ability to detect AI-generated text has yielded consistently mixed findings. \cite{Clark2021} found that even trained annotators performed only marginally above chance when distinguishing AI outputs from human writing, particularly for shorter texts. Importantly, \cite{Kbis2021} found that participants had difficulty reliably detecting AI-generated texts in incentivized Turing tests, particularly when a human selected the best AI outputs—though explicit warnings about potential AI authorship did not significantly improve detection accuracy but did reduce overall trust in content.\cite{Jakesch2023, Porter2024} demonstrated that people rely on flawed heuristics when judging AI-generated text, with AI systems capable of producing content perceived as “more human than human”.

A critical gap in this literature concerns whether detection ability relates to vulnerability to misinformation. Our study contributes to the literature by examining whether individual differences in IH predict detection accuracy and whether detection accuracy modulates susceptibility to AI-generated misinformation. We observed that the dissociation between source attribution and content evaluation suggests that these processes are functionally independent. If identifying content as AI-generated does not affect how individuals evaluate its credibility, then educational efforts focused on detection may be insufficient without accompanying training in content-based critical evaluation.

\subsection{Evolutionary Approaches to Epistemic Vigilance}

The concept of Error Management Theory \cite{Haselton2000} provides a complementary evolutionary framework, proposing that cognitive biases evolve when the costs of different types of errors (false positives vs. false negatives) are asymmetric over evolutionary time. \cite{Sperber2010} proposed that humans possess specific evolved mechanisms for epistemic vigilance, i.e., calibrated trust in communicated information based on assessments of both source reliability and content plausibility. This framework has been extended by \cite{Mercier2017}, who argued in The Enigma of Reason that human reasoning evolved primarily for social argumentation rather than individual cognition, with vigilance mechanisms serving to evaluate arguments presented by others.

We extend this evolutionary framework by proposing that IH may function as a threshold-setting mechanism for epistemic vigilance processes. Individuals higher in humility, by virtue of recognizing their own epistemic limitations, may maintain lower thresholds for triggering skeptical evaluation of incoming claims, but, crucially, this enhanced vigilance appears calibrated to detect exploitation attempts rather than to induce wide rejection of all novel information. Our interpretation has implications for understanding how foundation models might inadvertently exploit or fail to trigger ancestral defenses against unreliable information. AI-generated content, being evolutionarily unprecedented, may not activate source-based vigilance mechanisms that evolved to track human communicator reliability, while potentially evading content-based defenses through sophisticated mimicry of legitimate scientific discourse.

\section{Methods}

\subsection{Participants and Design}

99 university students from Maria Curie-Skłodowska University (UMCS) in Lublin, Poland, participated in the study (62 women, 33 men, 4 identifying as other gender; all aged 18-29 years). Participants were recruited through university communication channels and received no compensation. The sample size was determined by practical constraints in the educational context; sensitivity analysis indicated adequate power (1-$\beta$ > .80) to detect medium effect sizes (d = 0.50) for between-group comparisons.

Participants were randomly assigned to one of three conditions in a between-subjects design (n = 33 per condition), ensuring balanced group sizes. Each evaluated a single dialogue depicting a conversation between a user seeking advice about physical exercise for mental health, and an AI assistant providing recommendations. The three experimental conditions varied systematically in the scientific accuracy:

Condition 1 (Scientifically Accurate): The dialogue presented evidence-based information, citing established mechanisms such as brain-derived neurotrophic factor (BDNF) upregulation, cortisol regulation via the hypothalamic-pituitary-adrenal axis, modulation of serotonin and dopamine, and exercise-induced neuroplasticity. Claims were consistent with current neuroscientific literature on exercise and mental health.

Condition 2 (Moderately Pseudoscientific): The dialogue contained superficially plausible but scientifically unsupported claims, including assertions that exercise “detoxifies the brain”, that physical activity can serve as a complete replacement for psychotherapy in clinical depression, and vague references to “energy balancing” mechanisms without empirical grounding.

Condition 3 (Strongly Pseudoscientific): The dialogue invoked entirely fictional concepts, including “electromagnetic fields generated by contracting muscles that realign neural pathways”, “stress toxins” that accumulate in muscle tissue and are released through movement, and “cellular memory purification” processes with no basis in established human biology.

All dialogues were generated using a widely used LLM (GPT-4) to control for stylistic confounds, including writing quality, tone, vocabulary complexity, and conversational structure. The dialogues were matched in length and followed an identical conversational format: an initial user query, an AI response with recommendations, a follow-up user question, and a concluding AI response. This procedure ensured that any observed differences in credibility ratings could be attributed to content rather than surface-level textual features.

Following the dialogue evaluation, all participants received a comprehensive debriefing that explicitly identified which content was pseudoscientific, explained the scientific inaccuracies, and provided educational materials on evaluating the quality of health information. The debriefing referenced \cite{Bak2022} work on IH and included resources from the Polish Academy of Sciences on recognizing health misinformation.

\subsection{Measures}

Nine items assessed domain-specific IH regarding health knowledge, adapted from established IH measures \cite{KrumreiMancuso2015, Leary2017} to focus specifically on health-related beliefs. Items were rated on a 5-point Likert scale (1 = strongly disagree to 5 = strongly agree). The scale captured four conceptual dimensions: (a) awareness of knowledge limitations (e.g., “My views about health could be just as wrong as anyone else's”, “I don't know everything about health”), (b) openness to revising viewpoints (e.g., “I am open to new information that may change my views on health”), (c) recognition of information source limitations (e.g., “My sources of health information may not be the best”), and (d) acceptance of potential fallibility (e.g., “My views on health may turn out to be incorrect”). Internal consistency was acceptable (Cronbach's $\alpha$ = .79). Confirmatory analysis revealed substantial left-skewness across items, with ceiling effects particularly pronounced for items assessing openness to new information (Q6: M = 4.63, SD = 0.78) and willingness to change views given evidence (Q9: M = 4.61, SD = 0.74). 
Participants rated the dialogue on five dimensions using 7-point scales (1 = not at all to 7 = very much): (a) perceived credibility of information presented (“How credible do you find the information in this dialogue?”), (b) personal agreement with recommendations (“To what extent do you agree with the advice given?”), (c) likelihood of sharing with others (“How likely would you be to share this dialogue with someone seeking similar advice?”), (d) confidence in dialogue authenticity (“How confident are you that this represents a genuine conversation?”), and (e) trust in the information source (“How much do you trust the source providing this information?”). 
A categorical question assessed participants' beliefs about dialogue authorship: “Who do you believe created this dialogue?”. Possible responses were: (a) a conversation between two humans, (b) a conversation involving human and AI, or (c) uncertain/cannot determine. This question was positioned after credibility ratings to prevent source attribution from biasing quality judgments. An additional question assessed confirmation of bias tendencies: “When seeking health information, do you prefer sources that confirm your existing beliefs or sources that challenge them?”. It was an exploratory measure aimed to capture individual differences in epistemic motivation that might moderate responses to pseudoscientific content.

\subsection{Procedure}

The study was administered on-site via Google Forms. After providing informed consent, participants completed the IH scale, then read their assigned dialogue without time constraints, and immediately completed the dialogue evaluation measures, the source attribution question, and the information preference question. 

\subsection{Analysis}

Preliminary analyses revealed substantial violations of normality assumptions. Shapiro-Wilk tests indicated significant departures from normality for all nine IH items (all p < .001). Distributions exhibited pronounced negative skewness (range: -1.89 to -0.31) and variable kurtosis, reflecting ceiling effects characteristic of self-report humility measures. Consequently, we employed non-parametric statistical methods.

Between-group comparisons on dialogue evaluation measures used the Kruskal-Wallis H test, with post hoc pairwise comparisons conducted using the Mann-Whitney U test. To control for familywise error rate across the three pairwise comparisons per outcome variable, we applied Bonferroni correction (adjusted $\alpha$ = .0167). Effect sizes for Mann-Whitney tests were computed as r = Z/$\sqrt{N}$, interpreted following Cohen's conventions: small (r = .10), medium (r = .30), and large (r = .50). For chi-square analyses of categorical variables, effect sizes were reported as Cramér's V. 
Preliminary equivalence testing confirmed successful randomization: IH scores did not differ across conditions (Kruskal-Wallis H = 0.92, p = .631), nor did demographic distributions (gender: $\chi$² = 2.14, p = .343). All analyses were conducted in Python 3.11 using scipy.stats and pingouin packages, with significance set at $\alpha$ = .05.

\section{Results}

\subsection{Randomization and Distribution Checks}

Before testing the primary hypotheses, we verified successful randomization and examined the distributional properties of the key variables. IH scores did not differ across the three experimental conditions (Condition 1: M = 4.10, SD = 0.55; Condition 2: M = 4.10, SD = 0.52; Condition 3: M = 4.08, SD = 0.54; Kruskal-Wallis H = 0.92, p = .631), confirming that random assignment produced equivalent groups on this key moderator variable. Gender distributions were also equivalent across conditions ($\chi$² = 2.14, p = .343).

Examination of the distributions of the IH items revealed substantial ceiling effects, consistent with prior research on self-report humility measures. Items assessing openness to new information (Q6: M = 4.63, SD = 0.78) and willingness to revise beliefs given evidence (Q9: M = 4.61, SD = 0.74) showed the strongest ceiling effects, with over 70\% of participants endorsing the highest response option. In contrast, items requiring acknowledgment that one's own views might be as flawed as others' (Q1: M = 3.45, SD = 1.08) and that one's information sources may be suboptimal (Q5: M = 3.38, SD = 0.96) showed greater variability, suggesting these items capture more discriminating aspects of IH.

\subsection{Manipulation Check}

The experimental manipulation successfully produced differential credibility perceptions across conditions. A Kruskal-Wallis test revealed significant differences in perceived credibility (H = 26.61, df = 2, p < .001). Participants rated the scientifically accurate dialogue (M = 5.30, SD = 1.16, Mdn = 5.0) as substantially more credible than both pseudoscientific conditions.
Post-hoc Mann-Whitney U tests with Bonferroni correction (adjusted $\alpha$ = .0167) confirmed significant pairwise differences between the accurate condition and each pseudoscientific condition. The accurate dialogue was rated significantly higher than the moderately pseudoscientific dialogue (M = 3.73, SD = 1.35, Mdn = 4.0; U = 853.5, Z = 3.96, p < .001, r = .49, 95\% CI [.28, .65]), representing a medium-to-large effect. The accurate dialogue was also rated significantly higher than the strongly pseudoscientific dialogue (M = 3.36, SD = 1.54, Mdn = 3.0; U = 906.0, Z = 4.63, p < .001, r = .57, 9\% CI [.38, .72]), representing a large effect.
Critically, the two pseudoscientific conditions did not differ significantly from each other (U = 631.5, Z = 0.94, p = .349, r = .12). This null finding is theoretically important. Although the strongly pseudoscientific condition contained objectively more implausible claims (e.g., electromagnetic fields generated by muscles), participants did not differentiate between moderate and extreme pseudoscience. This pattern suggests a categorical rather than graded detection process, i.e., participants did not calibrate their skepticism to the severity of scientific violations.

The pattern of results was similar for personal agreement ratings (H = 20.26, p < .001), with participants agreeing more with accurate content (M = 5.64, SD = 1.06) than moderately (M = 4.09, SD = 1.47, r = .49) or strongly pseudoscientific (M = 4.03, SD = 1.57, r = .45). Trust ratings also differed across conditions (H = 6.34, p = .042), though pairwise comparisons revealed only the accurate-versus-moderate contrast reached significance after correction (r = .30). Notably, sharing intentions did not differ significantly across conditions (H = 3.30, p = .192), suggesting willingness to disseminate information was less sensitive to content quality than personal credibility judgments.

\subsection{Intellectual Humility as Selective Filter}

The central finding concerns the relationship between IH and credibility judgments and how it varies by content type. In the aggregate sample, IH was negatively correlated with credibility ratings (Spearman $\rho$ = -.231, p = .021; Kendall $\tau$ = -.177, p = .020; Pearson r = -.250, p = .013). However, such an overall correlation masks a theoretically crucial dissociation revealed by condition-specific analyses. 

Within the scientifically accurate condition, IH showed no relationship with credibility ratings ($\rho$ = -.038, p = .832, 95\% CI [-.38, .31]). The near-zero correlation and wide confidence interval spanning zero indicate that humble individuals were neither more nor less receptive to evidence-based health information than their less humble counterparts.

In stark contrast, IH was substantially and significantly correlated with credibility ratings in both pseudoscientific conditions. For moderately pseudoscientific content, the correlation was moderate-to-strong ($\rho$ = -.391, p = .024, 95\% CI [-.64, -.06]). For strongly pseudoscientific content, the correlation was even larger ($rho$ = -.462, p = .007, 95\% CI [-.69, -.14]). This suggests that individuals higher in IH assigned lower credibility to pseudoscientific claims, with the effect more pronounced for the most extreme misinformation.

\begin{figure}
    \centering
    \includegraphics[width=1\linewidth]{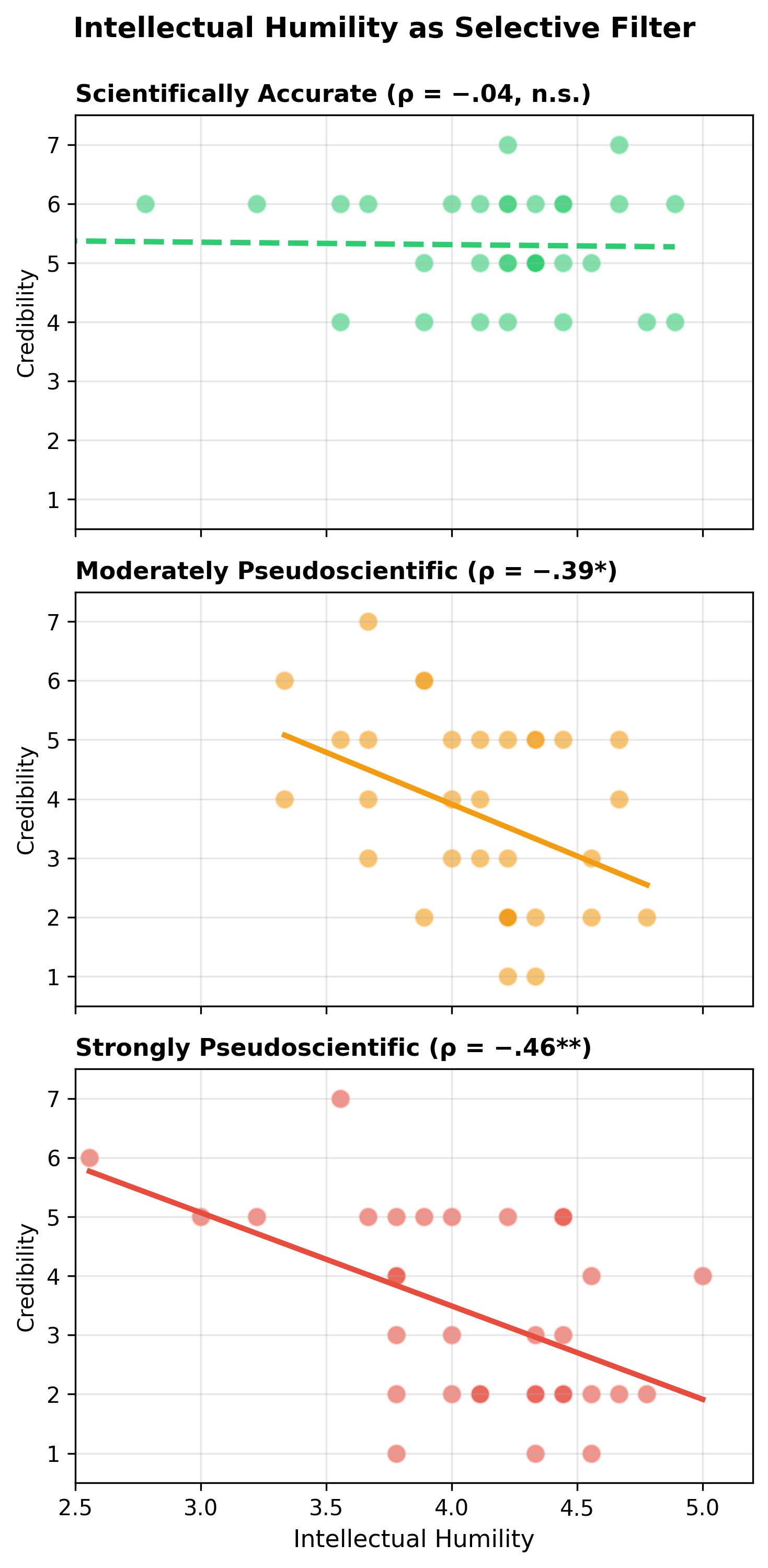}
    \caption{Intellectual Humility as Selective Filter}
\end{figure}

The revealed pattern of selective correlation provides evidence against two alternative interpretations. Firstly, IH does not induce generalized skepticism that would impair acceptance of all information regardless of quality. If humility operated as a nonspecific skepticism factor, we would expect negative correlations with credibility ratings across all conditions, including accurate content. Secondly, the condition-specific pattern cannot be attributed to the restriction of range in credibility ratings for accurate content. Although means were higher, variance was comparable across conditions (SD range: 1.16-1.54).

Instead, the findings support a calibrated vigilance interpretation. Intellectually humble individuals appear to maintain appropriately differentiated responses to information quality. They remain receptive to evidence-based claims while applying heightened scrutiny to pseudoscientific content. This selectivity is consistent with theoretical accounts proposing that IH facilitates accurate metacognitive monitoring rather than general doubt.

\subsection{Source Attribution}

Despite all dialogues being generated by an AI system, participants showed variable accuracy in source attribution. Overall, 61.6\% correctly identified the dialogue as involving AI, 28.3\% incorrectly attributed it to a human-human conversation, and 10.1\% indicated uncertainty.

Attribution accuracy varied descriptively across conditions in a counterintuitive pattern. Participants viewing scientifically accurate content were most likely to correctly identify AI involvement (69.7\%), followed by those viewing strongly pseudoscientific content (66.7\%), with the lowest accuracy for moderately pseudoscientific content (48.5\%). However, this pattern did not reach statistical significance ($\chi$² = 4.76, df = 4, p = .313, V = .16). The higher detection rate for accurate content is consistent with lay intuitions that AI systems produce too perfect or formulaic responses, though this interpretation remains speculative given the non-significant omnibus test.

\begin{figure}
    \centering
    \includegraphics[width=1\linewidth]{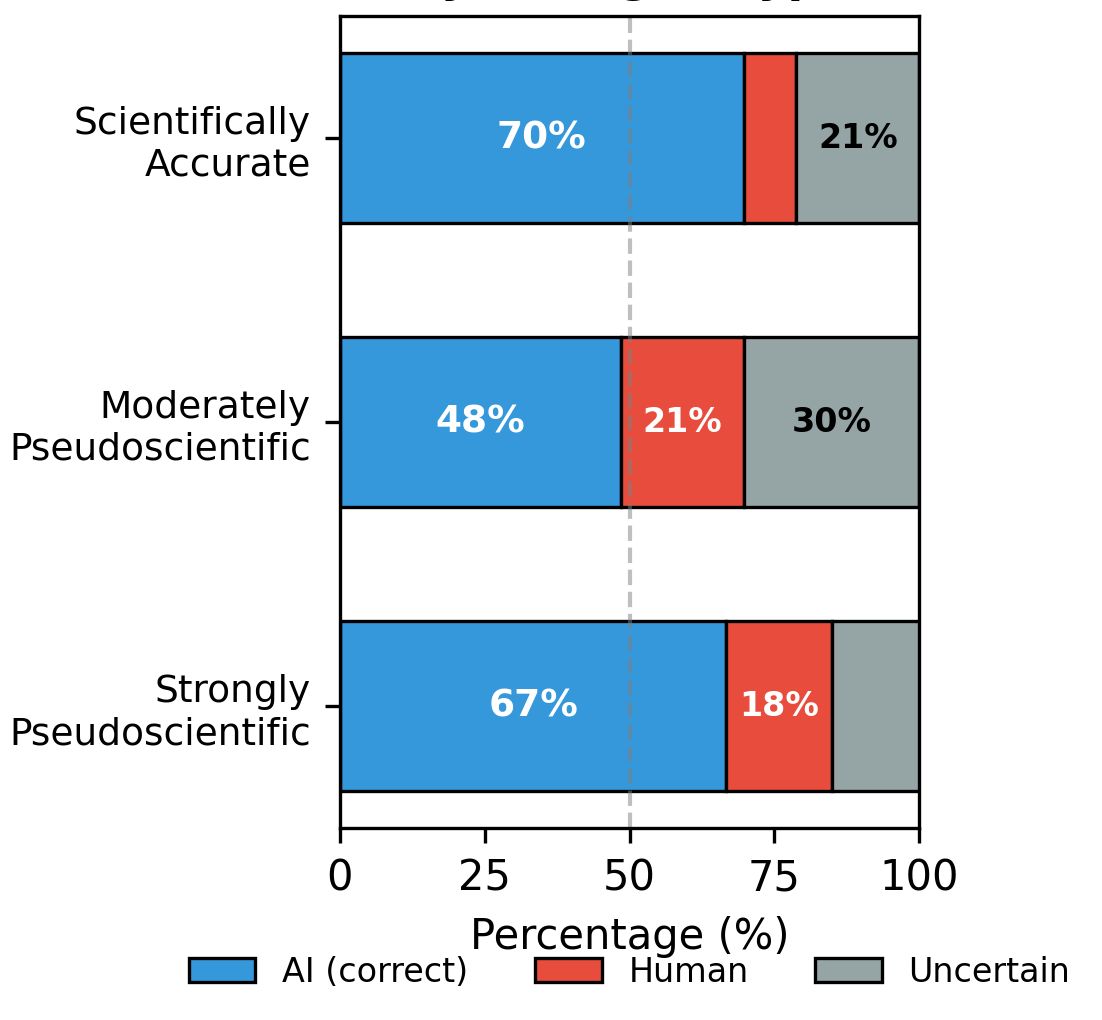}
    \caption{Source Attribution by Dialogue Type}
\end{figure}

IH did not predict source attribution accuracy ($\chi$² = 2.07, df = 4, p = .724, V = .10). Mean humility scores were virtually identical across attribution response categories (correct AI identification: M = 4.09; incorrect human attribution: M = 4.11; uncertain: M = 4.10; Kruskal-Wallis H = 0.12, p = .944). This null finding suggests that the cognitive processes underlying source attribution differ from those involved in content quality assessment, despite both potentially serving epistemic vigilance functions.

Critically, source attribution was not correlated with credibility judgments. Participants who correctly identified AI involvement did not differ in their credibility ratings from those who attributed the dialogue to humans or expressed uncertainty (Kruskal-Wallis H = 0.08, p = .959). Mean credibility ratings were nearly identical across categories (AI identified: M = 4.11; human attributed: M = 4.18; uncertain: M = 4.10). A chi-square test examining the association between attribution accuracy and credibility (dichotomized at the median) was also non-significant ($\chi$² = 0.94, df = 4, p = .919, V = .07). 

This double dissociation, i.e., humility predicting content evaluation but not source attribution, and source attribution being unrelated to content evaluation, has important theoretical and practical implications. It suggests that knowing (or believing) that content was AI-generated does not automatically trigger skepticism about content quality. The processes by which people evaluate what is said appear functionally independent from processes evaluating who (or what) said it.

\subsection{Confirmation Bias}

Approximately 60\% of participants across all conditions preferred information sources that confirm their existing beliefs to those that challenge them. This preference did not differ across experimental conditions ($\chi$² = 0.34, df = 2, p = .846), suggesting it reflects a stable individual difference rather than a response to experimental manipulation. The prevalence of confirmatory information preferences is consistent with well-documented confirmation bias tendencies in health information seeking and may partially explain why pseudoscientific content—which often aligns with intuitive but incorrect beliefs about health—can achieve acceptance even among otherwise skeptical individuals.

\subsection{Gender Differences}

Exploratory analyses revealed that women reported higher IH than men (women: M = 4.18, SD = 0.52; men: M = 3.93, SD = 0.54; U = 1329.5, Z = 2.40, p = .016, r = .25). This difference was driven primarily by items assessing acknowledgment that one's views could be wrong (Q1: r = .25, p = .012), recognition of knowledge limitations (Q3: r = .20, p = .017), and openness to new information (Q6: r = .17, p = .032). Given the modest effect size and exploratory nature, these findings should be interpreted cautiously pending replication.

\section{Discussion}

\subsection{Evolutionary Perspective}

We propose that IH functions as an ancestral adaptation for navigating informationally uncertain environments, i.e., one that remains adaptive for evaluating AI-generated content despite lacking evolutionary precedent for distinguishing artificial from human sources. From an error management perspective \cite{Haselton2000}, the costs of accepting false health information (potential harm) exceed the costs of rejecting true information (missed benefit), creating selection pressure for vigilance calibrated to content quality rather than source identity.

The dissociation between content evaluation and source attribution aligns with this interpretation. Throughout human evolutionary history, all communicated information originated mainly from humans. Discriminating between sources based on their artificial nature was never adaptively relevant. In contrast, detecting markers of unreliable or manipulative content (e.g., exaggerated promises, lack of uncertainty acknowledgment, appeals to mysterious mechanisms) could have consistently been beneficial across both ancestral and modern social environments.

\subsection{Implications for Foundation Models}

Our findings have implications for the responsible deployment of LLMs in health communication. Firstly, interventions targeting AI detection may be less effective than those strengthening content-based critical evaluation, as source attribution appears irrelevant to misinformation vulnerability. Secondly, foundation models generating health information should be designed to include epistemic markers that trigger vigilance in epistemically humble individuals, e.g., by acknowledging uncertainty, citing evidence, and avoiding superlative claims.

Paradoxically, our results suggest that accurate, well-calibrated AI outputs are more often identified as AI-generated than pseudoscientific content (69.7\% vs. 48.5\% in our data). This authenticity paradox may arise because precise, evidence-based language triggers associations with machine-like communication, while colloquial pseudoscience feels more human. It has concerning implications for trust in legitimate AI-assisted health communication.

\subsection{Observed and Declarative Humility}

The observed gap between declarative humility (high self-reported openness) and operational humility (preference for confirming evidence in 60\% of participants) raises questions for virtue epistemology. This divergence may reflect not hypocrisy but adaptive strategic signaling, e.g., publicly performing openness to maintain cooperative reputation while privately preserving belief coherence necessary for action. From this point of view, genuine IH may be rarer than survey measures suggest.

\section{Limitations}

Sample homogeneity (university students) limits generalizability. Older adults and those with lower education may show different patterns. The health domain focus may not extend to other misinformation contexts (e.g., political, financial). Future work should examine whether IH training can enhance pseudoscience detection and whether LLMs can be designed to appropriately engage human epistemic vigilance.

Direct Fisher r-to-z contrasts are underpowered at n = 33 per cell (accurate vs. strongly pseudoscientific, z = 1.79, p = .074). The selective-filter claim instead rests on a better-powered moderated regression pooling both pseudoscientific conditions against the accurate one, where the IH x content-type interaction was significant (b = -1.58, p = .002; R\textsuperscript{2} = .39).However, we regard this as suggestive rather than conclusive. Causal and functional readings are further limited by single-stimulus conditions, and unmeasured confounds (literacy, numeracy, education).

\section{Conclusions}

\subsection{Summary of findings}

Our findings demonstrate that IH functions as a selective cognitive filter that remains adaptive for evaluating AI-generated health information, discriminating between pseudoscientific and evidence-based content while operating independently of source-attribution processes. This selectivity, i.e., protecting against misinformation without inducing general skepticism toward accurate information, represents a sophisticated epistemic calibration mechanism that foundation models deployed for social good should respect and support. The dissociation between content evaluation and AI detection further underscores that users cannot be expected to protect themselves simply by identifying AI-generated content. The protective function of IH operates on what is said, not on attributions about who or what generated it. These findings carry direct implications for how AI systems should be designed, deployed, and regulated in health communication contexts.

\subsection{Design principles}

For developers of models intended for health applications, our results suggest several design principles. Firstly, AI systems should avoid mimicking the stylistic markers of pseudoscience, such as vague mechanistic claims, appeals to natural processes, and unfounded certainty, even when users prompt for such content, as intellectually humble users rely on these content-based cues for their epistemic filtering. Secondly, since users cannot distinguish moderate from extreme pseudoscience, AI guardrails should be calibrated conservatively. Once content crosses into unsupported claims, users lose the ability to grade its severity, making even mild misinformation potentially as damaging as obvious falsehoods. Thirdly, given that source attribution does not modulate credibility judgments, transparency labels indicating AI authorship, while valuable for other reasons, should not be relied upon as a primary safeguard against the acceptance of misinformation. Instead, foundation models should incorporate real-time epistemic scaffolding, i.e., explicitly flagging the evidence base for health claims, distinguishing established findings from emerging research, and proactively noting when user queries venture into domains where scientific consensus is lacking or where the AI's training data may be outdated. Such scaffolding would support rather than bypass the natural vigilance mechanisms our data reveal in intellectually humble users.

\subsection{Policy and implementation}

From a policy and implementation perspective, these findings argue for differentiated approaches to AI deployment across health contexts. Foundation models used in clinical decision support, public health messaging, and consumer health applications should undergo rigorous content validation that goes beyond factual accuracy to assess how claims might interact with user epistemic dispositions. Training programs for healthcare professionals integrating AI tools should emphasize that patients' ability to evaluate AI-generated recommendations depends on their epistemic traits, and not on whether they know the source is AI, suggesting that interventions promoting IH may be more protective than AI literacy campaigns focused solely on detection skills. Furthermore, regulatory frameworks such as the EU AI Act should consider content-level requirements alongside transparency mandates, recognizing that the harms from AI-generated health misinformation arise from the information itself rather than from users' failure to identify its AI origin. 

\subsection{The challenge of users' vulnerability}

Finally, our finding that approximately 60\% of users prefer confirmatory over challenging health information highlights a vulnerability that foundation models could either exploit or counteract. Socially beneficial AI should be designed to gently introduce corrective information and alternative perspectives, leveraging the openness to new evidence that characterizes intellectually humble users rather than reinforcing existing misconceptions through personalization algorithms optimized for engagement. Ultimately, the path toward foundation models that genuinely serve social good in health contexts requires designing systems that complement human epistemic strengths by amplifying the protective vigilance of IH, while compensating for documented weaknesses, including the categorical rather than graded nature of pseudoscience detection and the independence of content evaluation from source awareness.

\newpage

\bibliography{custom}

\end{document}